# Graphene-on-Diamond Devices with Enhanced Current-Carrying Capacity: Carbon *sp$^2$-on-sp$^3$* Technology

**Jie Yu[1,x], Guanxiong Liu[1,x], Anirudha V. Sumant[2,*], Vivek Goyal[1], and Alexander A. Balandin[1,*]**

[1]Nano-Device Laboratory, Department of Electrical Engineering and Materials Science and Engineering Program, University of California, Riverside, California 92521 USA

[2]Center for Nanoscale Materials, Argonne National Laboratory, Illinois, 60439 USA

[x]These authors contributed equally to this work

[*]Corresponding authors:

E-mail addresses: A.A.B. (balandin@ee.ucr.edu) and A.V.S. (sumant@anl.gov)

**Images for the Table of Context**

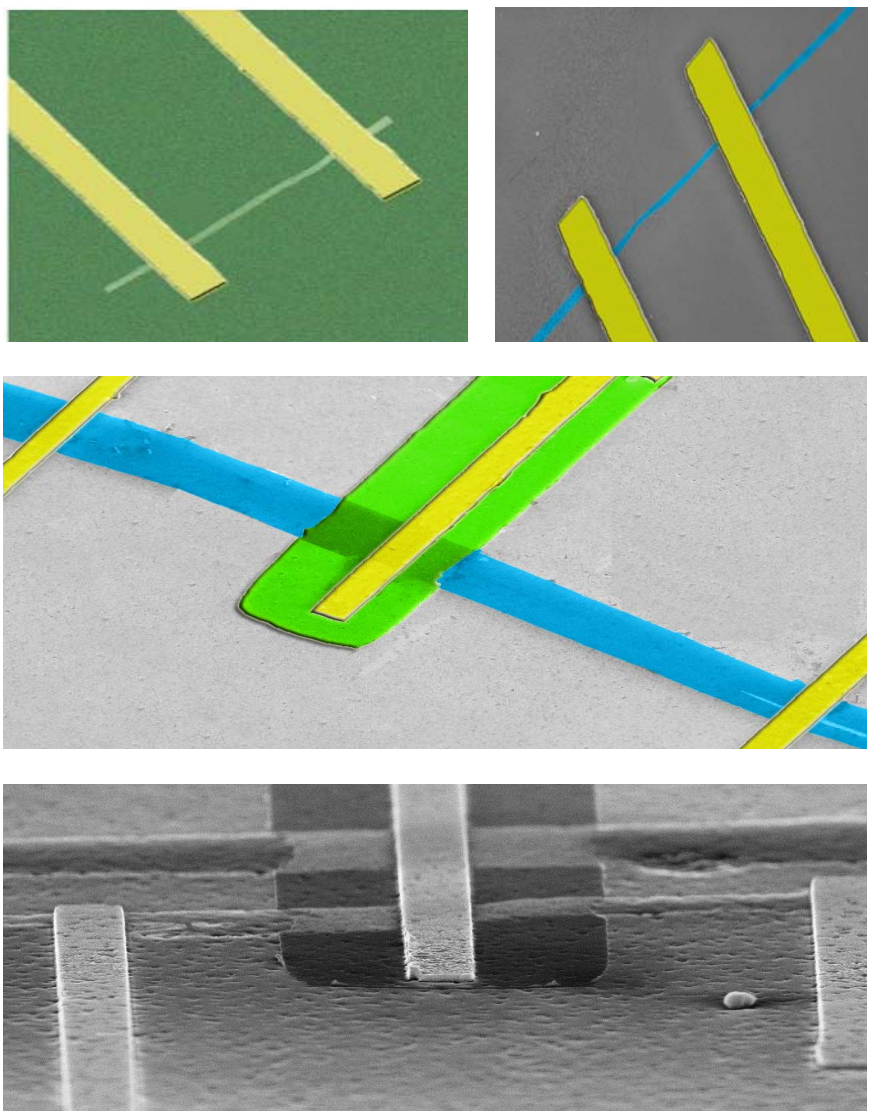

**Abstract**

Graphene demonstrated potential for practical applications owing to its excellent electronic and thermal properties. Typical graphene field-effect transistors and interconnects built on conventional $SiO_2$/Si substrates reveal the breakdown current density on the order of 1 $\mu A/nm^2$ (i.e. $10^8$ $A/cm^2$) which is ~100× larger than the fundamental limit for the metals but still smaller than the maximum achieved in carbon nanotubes. We show that by replacing $SiO_2$ with synthetic diamond one can substantially increase the current-carrying capacity of graphene to as high as ~18 $\mu A/nm^2$ even at ambient conditions. Our results indicate that graphene's current-induced breakdown is thermally activated. We also found that the current carrying capacity of graphene can be improved not only on the single-crystal diamond substrates but also on an inexpensive ultrananocrystalline diamond, which can be produced in a process compatible with a conventional Si technology. The latter was attributed to the decreased thermal resistance of the ultrananocrystalline diamond layer at elevated temperatures. The obtained results are important for graphene's applications in high-frequency transistors, interconnects, transparent electrodes and can lead to the new planar *sp$^2$-on-sp$^3$ carbon-on-carbon* technology.

Graphene is a promising material for future electronics owing to its high carrier mobility [1-2], saturation velocity [3], thermal conductivity [4-5], and ability to integrate with almost any substrate [6]. Particularly feasible are applications that do not require a bandgap but can capitalize on graphene's superior current-carrying capacity. Graphene field-effect transistors (FETs) and interconnects built on $SiO_2$/Si substrates reveal the breakdown current density, $J_{BR}$, of ~1 $\mu$A/nm$^2$ [7-9], which is ~100$\times$ larger than the fundamental electromigration limit for the metals [10]. However, the current-carrying capacity of graphene-on-$SiO_2$/Si devices typically reported in literature is still much smaller than the maximum achieved in carbon nanotubes (CNTs) [11-14]. Here, we used recent advances in the chemical vapor deposition (CVD) and processing of diamond for fabricating >40 graphene devices on ultrananocrystalline diamond (UNCD) and single-crystal diamond (SCD) substrates with the surface roughness below $\delta H \approx 1$ nm. It was found that not only SCD but also UNCD with the grain size $D$~5-10 nm can improve $J_{BR}$ owing to the increased thermal conductivity of UNCD at higher temperatures. The obtained results are important for graphene applications in interconnects [7, 15], radio-frequency (r.f.) transistors [16], and can lead to the new planar *sp$^2$-on-sp$^3$ carbon-on-carbon* technology with superior current-carrying capacity. A possibility of direct growth of graphene on diamond or graphitization of the top diamond layers for graphene device fabrication can provide another impetus to the planar sp$^2$-on-sp$^3$ carbon-on-carbon technology.

Graphene devices are commonly fabricated on Si/$SiO_2$ substrates with the $SiO_2$ thickness of $H \approx 300$ nm [1-3]. Owing to optical interference, graphene becomes visible on Si/$SiO_2$ (300-nm) substrates, which facilitates its identification. Graphene reveals excellent heat conduction properties with the intrinsic thermal conductivity, $K$, exceeding 2000 W/mK at room temperature (RT) [4-5]. However, in typical device structures, e.g., FETs or interconnects, most of heat propagates directly below the graphene channel in the direction of the heat sink, i.e. bottom of Si wafer [17-18]. For this reason, the highly thermally resistive $SiO_2$ layers act as the *thermal bottleneck*, not allowing one to capitalize on graphene's excellent intrinsic properties. Theory suggests that the breakdown mechanism in $sp^2$-bonded graphene should be similar to that in $sp^2$-bonded CNTs. Unlike in metals, the breakdown in CNTs was attributed to the resistive heating or local oxidation, assisted by defects [11-14]. Thermal conductivity of $SiO_2$ $K$=0.5 − 1.4 W/mK at RT [19], is more than 1000-times smaller than that of Si, $K$=145 W/mK, which suggests that the use of materials with higher $K$, directly below graphene, can improve graphene's $J_{BR}$, and reach the maximum values observed for CNTs.

Synthetic diamond is a natural candidate for the use as a bottom dielectric in graphene devices, which can perform a function of heat spreader. Recent years witness a major progress in CVD diamond growth performed at low temperature, $T$, compatible with Si complementary metal-oxide-semiconductor (CMOS) technology [20-22]. There are other potential benefits of diamond layers utilized instead of $SiO_2$ in the substrates for graphene devices. The energy of the optical phonons in diamond, $E_p$=165 meV, is much larger than that in $SiO_2$, $E_p$=59 meV. The latter can improve the saturation velocity in graphene when it is limited by the surface electron − phonon scattering [23]. The lower trap density achievable in diamond, compared to $SiO_2$, indicates a possibility of reduction of the $1/f$ noise in graphene-on-diamond devices [24], which is essential for applications in r.f. transistors and interconnects.

Recently, it was demonstrated that replacing $SiO_2$ with diamond-like carbon (DLC) helps one to substantially improve the r.f. characteristics of the graphene transistors [16]. However, DLC is an amorphous material with $K$=0.2 − 3.5 W/mK at RT [25], which is a very low value even compared to $SiO_2$. Depending on H content, as-deposited DLC films have high internal stress, which needs to be released by annealing at higher $T \sim 600$ ${}^o$C [26]. These facts provide strong motivations for the search of other carbon materials, which can be used as substrates for graphene devices.

Synthetic diamond can be grown in a variety of forms from UNCD films with the small grain size, $D$, and, correspondingly low $K$, to SCD, with the highest $K$ among all bulk solids. Microcrystalline diamond (MCD) has larger $D$ than that of UNCD but suffers from unacceptable surface roughness, $\delta H$, and high thermal boundary resistance, $R_B$ [5]. Up to date, despite attempts in many groups to fabricate graphene devices on diamond with acceptable characteristics, no breakthrough was reported. The major stumbling blocks for development of viable graphene-on-diamond $sp^2$-on-$sp^3$ technology are high $\delta H$ of synthetic diamond, difficulty of visualization of graphene on diamond and problems with the top-gate fabrication – no bottom gates are possible on SCD substrates. We used the most recent advances in CVD diamond growth and polishing as well as our experience of graphene device fabrication to prepare a large number of test-structures, and study the current-carrying and thermal characteristics of graphene-on-diamond devices in the practically relevant ambient conditions. We considered two main forms of diamond – UNCD and SCD – which represent two extreme cases, in terms of $D$ and $K$.

The UNCD films for this study were grown on Si substrates in the microwave plasma chemical vapor deposition (MPCVD) system at the Argonne National Laboratory (ANL). Figure 1a-b shows the MPCVD system used for the growth inside a cleanroom and schematic of the process, respectively. The growth conditions were altered to obtain larger $D$, in the range 5-10 nm, instead of typical grain sizes $D \approx 2\text{-}5$ nm in UNCDs. This was done to increase $K$ of UNCD without strongly increasing the surface roughness. We intentionally did not increase $D$ beyond 10 nm or used MCD in order to keep $\delta H$ in the range suitable for polishing. The inset shows a 100-mm UNCD/Si wafer.

[**Figure 1**]

The surface roughness of the synthetic diamond substrate plays an important role in reducing electron scattering at the graphene – diamond interface and increasing the electron mobility, $\mu$. We performed the chemical mechanical polishing (CMP) to reduce the as-grown surface roughness from $\delta H \approx 4$-7 nm to below $\delta H \approx 1$ nm, which resulted in a corresponding reduction of the thickness, $H$, from the as-grown $H \approx 1$ μm to ~700 nm. The $H$ value was selected keeping in mind conditions for graphene visualization on UNCD together with the thermal management requirements (see *Methods* and *Supporting Online Material*). The SCD substrates were type IIb (100) grown epitaxially on a seed diamond crystal and then laser cut from the seed. For graphene devices fabrication, the SCD substrates were acid washed, solvent cleaned and put through the hydrogen termination process [27]. The near-edge x-ray absorption fine-structure spectrum (NEXAFS) of the grown UNCD film confirms its high $sp^3$ content and quality (Figure 1c). The strong reduction of $\delta H$ is evident from the atomic force microscopy (AFM) images of the as-grown UNCD and UNCD after CMP presented in Figure 1d and 1e, respectively. Details of the original growth process developed at ANL and the surface treatment procedures used for this study are given in the *Supporting Online Material*.

For the prove-of-concept demonstration, graphene and few-layer graphene (FLG) were prepared by exfoliation from the bulk highly oriented pyrolytic graphite to ensure the highest quality and uniformity. We selected flakes of the rectangular-ribbon shape with the width $W{\geq}1$ μm, which is larger than the phonon mean free path $\Lambda{\sim}750$ nm in graphene [5]. The condition $W{>}\Lambda$ ensured that $K$ does not undergo additional degradation due to the phonon-edge scattering, allowing us to study the breakdown limit of graphene itself. The length, $L$, of graphene ribbons was in the range 10-60 μm. We further chose ribbons with the small aspect ratio $\gamma{=}W/L{\sim}0.03$-0.1 to imitate interconnects. One should note that the current-carrying ability of graphene would depend on the length of the graphene channel, its width, aspect ratio and quality.

Raman spectroscopy was used for determining the number of atomic planes, $n$, in FLG although the presence of $sp^2$ carbon at the grain boundaries in UNCD made the spectrum analysis more difficult. Figure 1f shows spectra of the graphene-on-UNCD/Si and UNCD/Si substrate. One can see $1332$ cm$^{-1}$ peak, which corresponds to the optical vibrations in the diamond crystal structure. The peak is broadened due to the small $D$ in UNCD. The bands at ~ 1170, 1500 and 1460 cm$^{-1}$ are associated with the presence of trans-poly-acetylene and $sp^2$ phase at grain boundaries [28-29]. The graphene $G$ peak at 1582 cm$^{-1}$ and $2D$ band at ~2700 cm$^{-1}$ are clearly recognizable. Figure 1g presents spectra of the graphene-on-SCD, SCD substrate and difference between the two. The intensity and width of 1332 cm$^{-1}$ peak confirms that we have single-crystal diamond. We used both comparison of the intensity of the $G$ and $2D$ peaks and deconvolution of 2D band for determining the number of layers. We had prior experience of determining n in FLG samples on various substrates [30-35]. One can see from Figure 1g that the Raman spectrum of graphene on diamond after subtraction of diamond Raman signal looks similar to that of graphene on the standard Si/SiO$_2$ wafer and can be readily used for the number of layers counting. For a number of selected samples we used AFM inspection to verify $n$. The results obtained with the micro-Raman and AFM techniques were in agreement, which confirmed that the Raman method is very accurate, particularly for $n \leq 5$.

We intentionally focused on devices made of FLG with $n \leq 5$. FLG supported on substrates or embedded between dielectrics preserves its transport properties better than single-layer graphene. Two-terminal (i.e. interconnects) and three-terminal (i.e. FETs) devices were fabricated on both UNCD/Si and SCD substrates. The electron-beam lithography (EBL) was used to define the source, drain contacts, and gate electrodes. The contacts consisted of a thin Ti film covered by a thicker Au film. The top-gate $HfO_2$ dielectric was grown by the atomic layer deposition (ALD). The novelty in our design, as compared to the graphene-on-$SiO_2$/Si devices, was the fact that the gate electrode and pad were completely separated by $HfO_2$ layer to avoid oxide lift-off sharp edges, which can affect connection of the gate electrode. Figure 2a shows schematics of the fabricated devices. For testing the breakdown current density in FLG we used two-terminal devices in order to minimize extrinsic effects on the current and heat conduction. Three-terminal devices were utilized for $\mu$ measurements. We also fabricated conventional graphene-on-$SiO_2$/Si devices as references. Figure 2b is an optical microscopy image of two-terminal graphene-on-SCD devices. Figures 2c and 2d show the scanning electron microscopy (SEM) images of the two-terminal and three-terminal graphene-on-UNCD devices, respectively.

[**Figure 2**]

We electrically characterised >40 graphene-on-diamond devices and >10 graphene-on-$SiO_2$/Si reference devices. To understand the origin of the breakdown we correlated $J_{BR}$ values with the thermal resistances of the substrates. We measured the *effective K* of the substrates and determined their thermal resistance as $R_T = H_S/K$, where $H_S$ is the substrate thickness. For details of the thermal measurements see the *Supporting Online Material*. Figure 3a shows $R_T$ for the UNCD/Si and Si/SiO$_2$ (300-nm) substrates as a function of $T$. Note that $R_T$ for Si increases approximately linear with $T$, which is expected because the intrinsic thermal conductivity of crystalline materials decreases as $K \sim 1/T$ for $T$ above RT. The $T$ dependence of $R_T$ for UNCD/Si is notably different, which results from interplay of heat conduction in UNCD and Si. In UNCD, $K$ grows with temperature owing to increasing inter-grain transparency for the acoustic phonons that carry heat [5]. UNCD/Si substrates, despite being more thermally resistive than Si wafers at RT, can become less thermally resistive at high $T$. The $R_T$ value for SCD substrate is $\sim 0.25 \times 10^{-6}$ m$^2$K/W, which is more than order-of-magnitude smaller than that of Si at RT. The thermal interface resistance, $R_B$, between FLG and the substrates is $R_B \approx 10^{-8}$ m$^2$K/W, and it does not strongly depend on either $n$ or the substrate material [5]. For this reason, $R_B$ does not affect the $R_T$ trends.

Figure 3b shows current-voltage (I-V) characteristics of graphene-on-SCD FET at low source-drain voltages for different top-gate, $V_{TG}$, bias. The inset demonstrates a high quality of the HfO$_2$ dielectric and metal gate deposited on top of graphene channel. The linearity of I-Vs confirms that the contacts are Ohmic. Figure 3c presents the source-drain, $I_{SD}$, current as a function of $V_{TG}$ for graphene-on-UNCD FET. In the good top-gate graphene-on-diamond devices the extracted $\mu$ was $\sim 1520$ cm$^2$V$^{-1}$s$^{-1}$ for electrons and $\sim 2590$ cm$^2$V$^{-1}$s$^{-1}$ for holes. These mobility values are acceptable for applications in downscaled electronics. In Figure 3d we show results of the breakdown testing at ambient conditions. For graphene-on-UNCD, we obtained $J_{BR} \approx 5 \times 10^8$ A/cm$^2$ as the highest value, while the majority of devices broke at $J_{BR} \approx 2 \times 10^8$ A/cm$^2$. The reference graphene-on-SiO$_2$/Si had $J_{BR} \approx 10^8$ A/cm$^2$, which is consistent with literature [7-9]. The comparison with our own reference devices is more meaningful because they had similar graphene channel length, width, aspect ratio, quality and location of the metal pads, which serve as additional heat sinks.

The maximum achieved for graphene-on-SCD was as high as $J_{BR} \approx 1.8 \times 10^9$ A/cm$^2$. This is an important result, which shows that via improved heat removal from graphene channel one can reach, and even exceed, the maximum current-carrying capacity of ~10 µA/nm$^2$ (=1×10$^9$ A/cm$^2$) reported for CNTs [11-14]. The surprising improvement in $J_{BR}$ for graphene-on-UNCD is explained by the reduced $R_T$ of the substrate at high $T$ where the thermally-activated failure occurs. At this temperature, $R_T$ of UNCD/Si can be lower than that of Si/SiO$_2$ (see Figure 3a).

It is illustrative to perform a detailed comparison of $J_{BR}$ for graphene-on diamond devices with some recently reported results for graphene-on-SiO$_2$/Si. It was recently found that the breakdown current-density can be increased in graphene nanoribbons to the value of 4×10$^8$ A/cm$^2$ [36]. The highest value was found for one sample that had the smallest width of 15 nm of all examined graphene ribbons. The authors attributed this increase specifically to the extremely narrow width of the ribbons and provided heat dissipation argument. Heat spreading from the narrow ribbon will be three-dimensional, i.e. in all directions, while in graphene devices with wider ribbons the heat spreading will be mostly in vertical direction down to the heat sink [36]. Although the value for nanoribbon cannot be used for direct comparison with our data for graphene channels with the few-micrometer width ($J_{BR} \sim 10^8$ A/cm$^2$ for our reference graphene-on-SiO$_2$/Si devices), one can conclude that graphene devices on diamond with the nanometer width can have even larger breakdown current density.

The short length of the graphene channels and proximity of the metal pads, which serve as additional heat sinks can also affect the breakdown current density. It was reported that a graphene sample with the width of 4 μm and length of only 1 μm connected to the large source and drains had the breakdown current density of $3 \times 10^8$ A/cm$^2$ [37]. The samples studied in our work had the "opposite" interconnect-like geometry – the length was much larger than the width, i.e. $\gamma = W/L \sim 0.03$-$0.1$, although both $W$ and $L$ were in the micrometer range. The latter suggests that in some applications, where $\gamma \sim 1$ and $L \sim 1$ μm, the current-carrying ability of the graphene-on-diamond devices can be further increased owing to closer location of the metal heat sinks and better lateral heat spreading. We note here that FLG ribbons studied in Ref. [7] had the width of 22 nm and length of 0.75 μm. For this reason, the close proximity of the source and drain metal contacts could have influenced the overall value of the breakdown current density. One should also mention that in our experiments with graphene-on-diamond and reference graphene-on-SiO$_2$/Si the breakdown was achieved before clear signatures of the drain current saturation. The values of the breakdown current can be different in graphene nanoribbon transistors with the nanometer channel length in the current saturation regime [38-39].

[**Figure 3**]

The location of the current-induced failure spot and $J_{BR}$ dependence on electrical resistivity, $\rho$, and length, $L$, can shed light on the physical mechanism of the breakdown. The failures in the middle of CNTs and $J_{BR} \sim 1/\rho$ were interpreted as signatures of the electron diffusive transport, which resulted in the highest Joule heating in the middle [11-13]. The failures at the CNT-metal contact were attributed to the electron ballistic transport through CNT and energy release at the contact. There is a difference in contacting CNT with the diameter $d \sim 1$ nm and graphene ribbons with $W \geq 1$ μm. It is easier to break CNT-metal than the graphene-metal contact thermally. In our study, we observed the failures both in the middle and near the contact regions. Figure 4 shows $J_{BR}$ data for both types of the breakdown occurring in graphene-on-UNCD samples with similar aspect ratio $\gamma$. The difference between these two types of the breakdown was less pronounced in our graphene samples than that in CNTs. The failures occurred not exactly at the graphene-metal interface but on some distance, which varied from sample to sample. We attributed it to the width variations in graphene ribbons leading to breakdowns in the narrowest regions, or in the regions with defects, which are distributed randomly.

[**Figure 4**]

We did not observe scaling of $J_{BR}$ with $\rho$ like in the case of CNTs. Intriguingly, $J_{BR}$ for graphene scaled well with $\rho L$. From the fit to the experimental data we obtained $J_{BR} = \alpha (\rho L)^{-\beta}$, where $\alpha = 1.3 \times 10^{-6}$ and $\beta = 0.73$ for graphene-on-UNCD. For graphene-on-SCD, the slope is $\beta = 0.51$. Previously, the scaling with $(\rho L)^{-\beta}$ (where $\beta = 0.6$-$0.7$) was observed in carbon nanofibers (CNF) [40], which had a similar aspect ratio. Such $J_{BR}(\rho L)$ dependence was explained from the solution of the heat-diffusion equation, which included thermal coupling to the substrate. However, the thermally-induced $J_{BR}$ for CNF was $\sim 10^6$ A/cm² – much smaller than the record $J_{BR} \approx 1.8 \times 10^9$ A/cm² we obtained for graphene-on-SCD. All our measurements have been performed under ambient conditions where the thermal breakdown can be facilitated by oxidation. The oxidation temperature is likely in the range 600 – 800 °C [7, 36]. One should expect that the JBR for the high quality graphene-on-diamond in vacuum will be substantially higher.

In conclusion, we demonstrated, in a systematic study, that replacing $SiO_2$ with synthetic diamond allows one to achieve graphene's intrinsic current-carrying capacitance limit, which is on the same order of magnitude as that in carbon nanotubes. We confirmed that graphene's current-induced breakdown is thermally activated. It was also found that inexpensive UNCD/Si substrates, which are produced at CMOS compatible temperatures, can be used for improving the breakdown current density in graphene devices. The measured maximum breakdown current density $J_{BR}$ in ambient for graphene-on-UNCD and graphene-on-SCD was $5 \times 10^8$ A/cm$^2$ and $18 \times 10^8$ A/cm$^2$, respectively. For comparison, the reference graphene-on-$SiO_2$/Si samples, which had similar geometry of the graphene channel and identical heat sinks, had the breakdown current density of $1 \times 10^8$ A/cm$^2$. Our results together with the prospects of direct growth of graphene on diamond or graphitization of the top diamond layers for graphene device fabrication can stimulate development of the planar sp²-on-sp³ carbon-on-carbon technology.

**METHODS**

The UNCD thin films were grown on 100-mm diameter Si substrates in 915 MHz large-area microwave plasma chemical vapor deposition (MPCVD) system (DiamoTek 1800 series 915 MHz, 10 KW from Lambda Technologies Inc.) operating in the clean room at the Argonne National Laboratory. Prior to the growth, silicon substrate were deposited with 10 nm tungsten layer using sputter deposition process followed by nanodiamond seeding treatment using the nanodiamond suspension containing dimethylsulphoxide (DMSO) solution (ITC, Raleigh, NC). Details about MPCVD and seeding process for the UNCD growth are described in the *Supplementary Information*. The single crystal diamonds used for this study were type IIb with (100) orientation (Delaware Diamond Knives) polished from both sides down to ~3-nm RMS roughness. A pre-cleaning procedure using acid wash and solvent cleaning was used to etch any contaminants from the surface. The H-termination process with microwave plasma was carried at the substrate $T$=700 $^o$C using $H_2$ flow of 50 sccm and chamber pressure of 30 mbar for 10-15 mins. The process eliminates any hydrocarbon and oxygenated impurities and produces clean H-terminated diamond surface. We defined the top-gate region using EBL (NPGS controlled Leo 1550) and performed ALD (Cambridge Nanotech) of 20-nm thick $HfO_2$ at $T$=110°C. The lift-off of ALD was done in hot acetone ($T$=60°C) for ~2 hours. We often observed oxide leftovers at the edges of the defined regions, which can lead to discontinuities in the following metal layer. To avoid this problem, we designed $HfO_2$-layer insert under the entire region of gate electrode and pad. We then used EBL to define the source, drain and top gate electrodes regions and deposit Ti/Au (10nm/100nm) by E-beam evaporator (Temescal BJD-1800).The gate leakage in our devices was very low (much smaller than 0.1 nA/$\mu$m$^2$). We established that our polished UNCD/Si substrates do not require a seeding layer for ALD of $HfO_2$ gate dielectric. The near edge x-ray absorption fine structure spectroscopy (NEXAFS) of UNCD sample was carried out at the University of Wisconsin Synchrotron Radiation Center Facility. The data was acquired at HERMON beam at carbon K edge with high energy resolution (0.2-0.4 eV). The spectra were taken in the total electron yield (TEY) mode with the incident photon beam normal to the sample. Special care was taken to correct for the carbon contamination from the x-ray beam optics and transmission structure from the monochromator. Details of the measurements are described in the *Supplemental Information*.

### *Acknowledgements*

The work at UCR was supported by the Office of Naval Research (ONR) through award N00014-10-1-0224, Semiconductor Research Corporation (SRC) and Defense Advanced Research Project Agency (DARPA) through FCRP Center on Functional Engineered Nano Architectonics (FENA), and DARPA Defense Microelectronics Activity (DMEA) under agreement number H94003-10-2-1003. The work at ANL was supported by the U.S. Department of Energy (DOE), Office of Science and Office of Basic Energy Sciences under Contract DE-AC02-06CH11357. NEXAFS studies were performed at the University of Wisconsin Synchrotron Radiation Center.

**Author Contributions** A.A.B. coordinated the project, led the graphene device data analysis and wrote the manuscript; A.V.S. developed diamond growth and polishing processes, performed diamond characterization, contributed to data analysis and manuscript writing; J.Y. performed graphene device fabrication, Raman and electrical measurements; G.L. performed graphene device fabrication and contributed to data analysis; V.G. performed thermal measurements.

**Author Information** Correspondence and requests for materials should be addressed to A.A.B. (balandin@ee.ucr.edu) and A.V.S. (sumant@anl.gov).

**Present Address Information:** V.G. is with Texas Instruments, Dallas, Texas 75243 USA

**Supporting Information Available:** Description of the material preparation and details of the thermal measurements. This material is available free of charge via the Internet at http://pubs.acs.org

**Figure Captions**

**Figure 1:** (A): Large-area MPCVD system used for the synthetic diamond growth. The inset shows a 100-mm Si/UNCD wafer. (B): Schematics describing the UNCD growth in the MPCVD system. (C): NEXAFS data for deposited UNCD thin film revealing its high $sp^3$ content and quality. The exciton peak at ~289.3 eV corresponds to 1s→ σ* resonance from $sp^3$ carbon. The peak at ~285 eV corresponds to 1s→ π* resonance from sp² carbon at grain boundaries. The revealed $sp^2$ fraction is 2%, which is lower than the typical 5% $sp^2$ content, owing to larger $D$ in our UNCD. (D) and (E): AFM images of the as-grown and chemical-mechanical polished UNCD, respectively. (F): Raman spectra of graphene-on-UNCD and UNCD substrate. (G): Raman spectra of graphene-on-SCD and SCD substrate. The difference in spectra was used to determine the number of atomic planes, $n$. The specific example shows single-layer graphene.

**Figure 2:** (A): Schematic of the two-terminal and three-terminal devices fabricated for testing on UNCD/Si and SCD substrates. (B): Optical microscopy image of the two-terminal graphene devices – prototype interconnects – on single-crystal synthetic diamond. (C) and (D): SEM images of the two-terminal and three-terminal graphene-on-UNCD/Si devices. The two-terminal devices were used for the breakdown current testing, while the three-terminal devices – for the mobility. The scale bar is 2 μm.

**Figure 3:** (A): Thermal resistance of UNCD/Si substrate and reference Si wafer. (B): Low-field current-voltage characteristics $I_{DS}$ vs. $V_{TG}$ for the top-gate graphene-on-SCD devices. The top

gate bias $V_{TG}$ varies from -4.0 V to +4.0 V with the step of 1.0 V. (C): Source-drain current in the three-terminal graphene-on-UNCD devices vs. the top-gate bias. (D): Breakdown current density in the two-terminal graphene-on-UNCD and graphene-on-SCD devices. Note an order of magnitude improvement in the current-carrying ability of graphene devices fabricated on single-crystal synthetic diamond.

**Figure 4:** Breakdown current density $J_{BR}$ as a function of the electrical resistance and length of graphene interconnects. The device failures close to the middle of the graphene channel and to the graphene - metal contact are indicated with red circles and blue rectangulars, respectively.

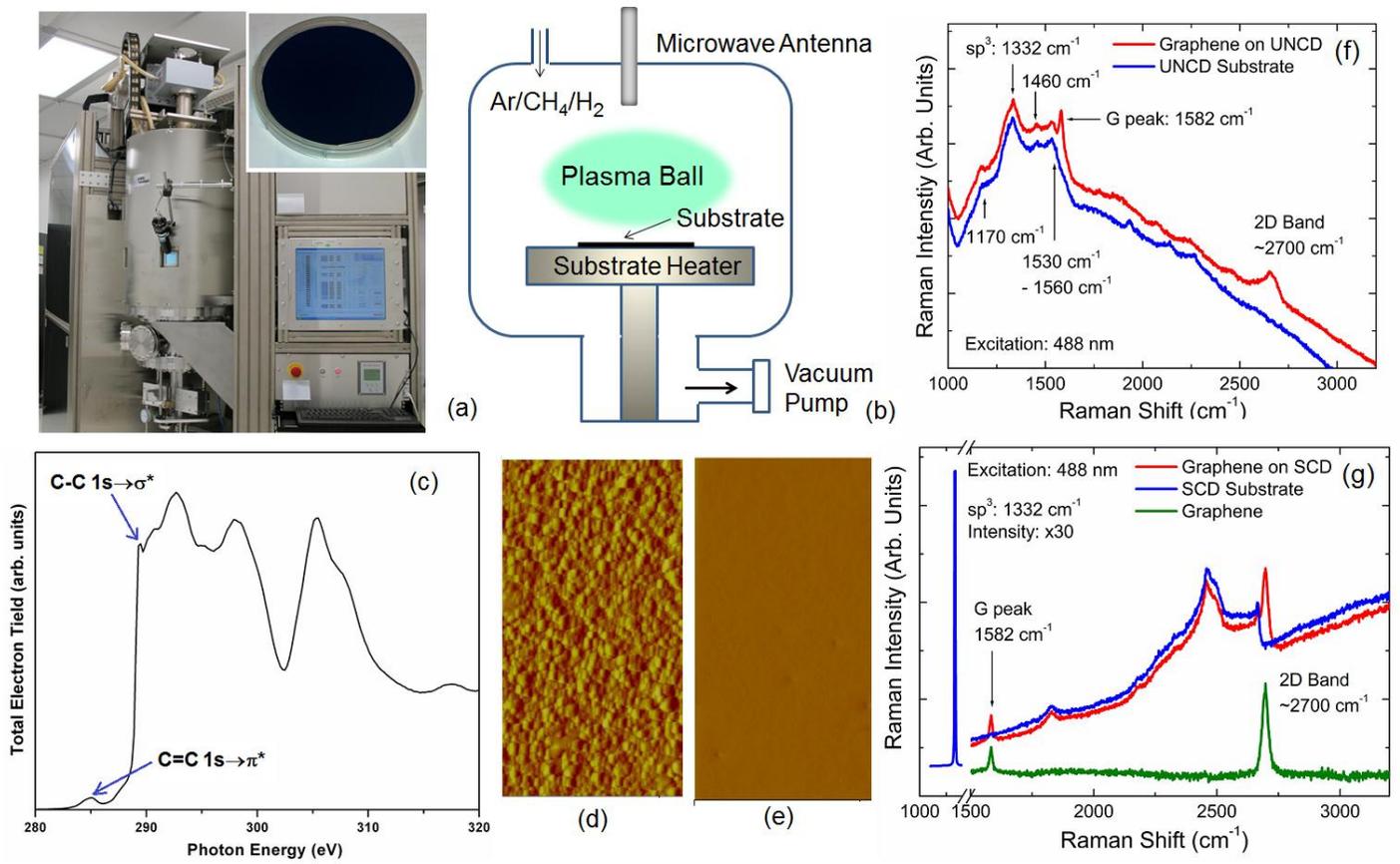

Figure 1

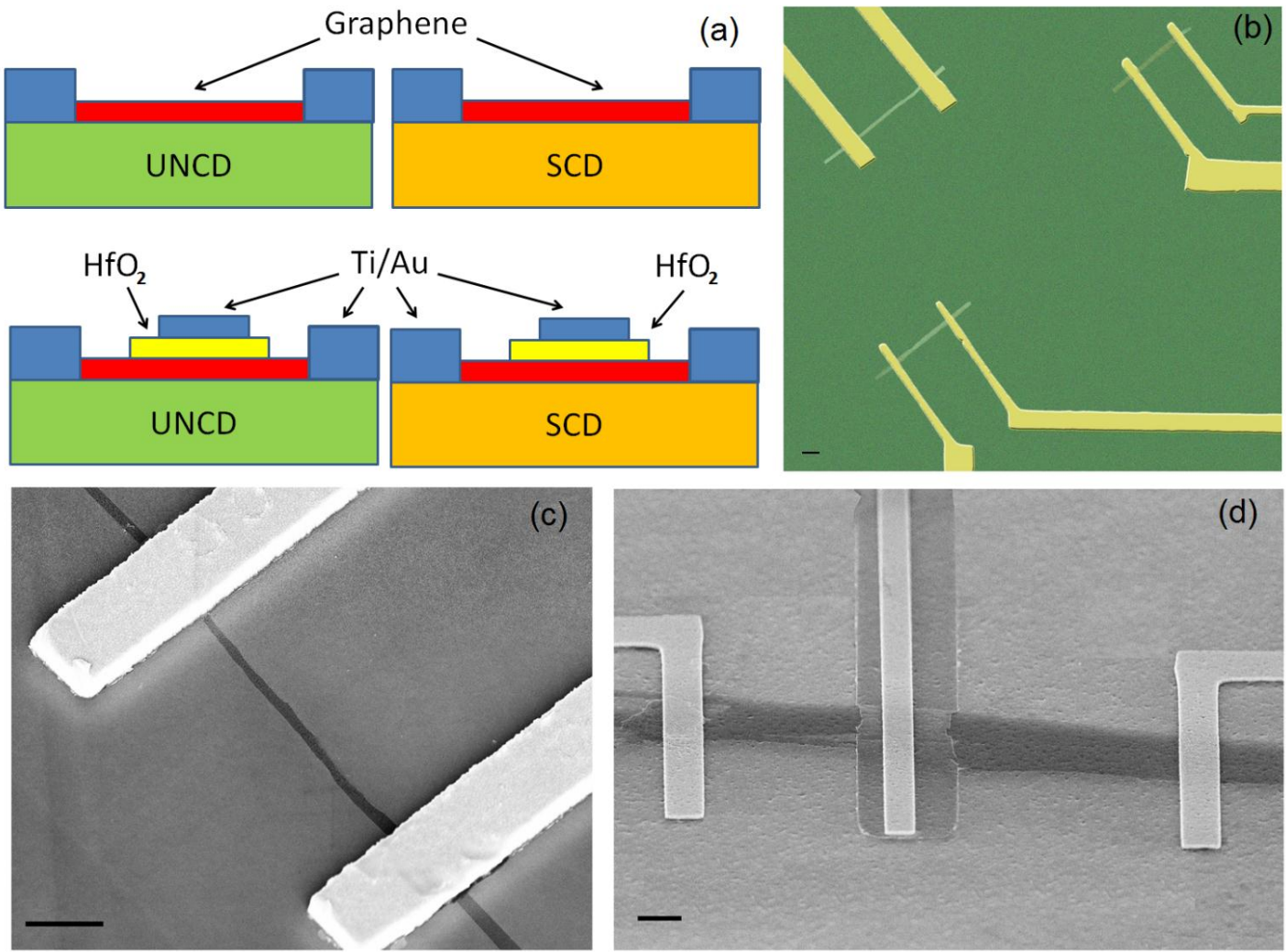

Figure 2

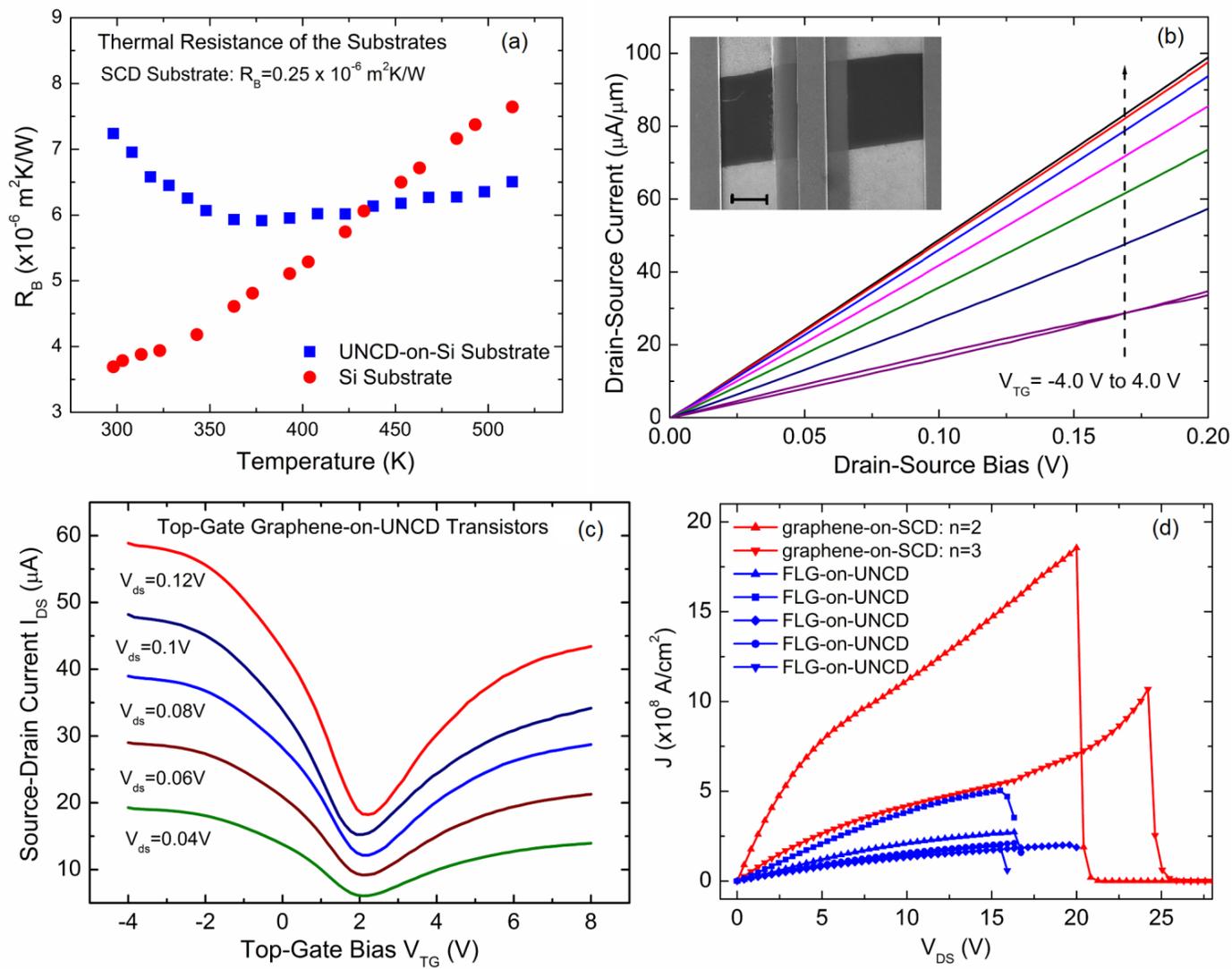

Figure 3

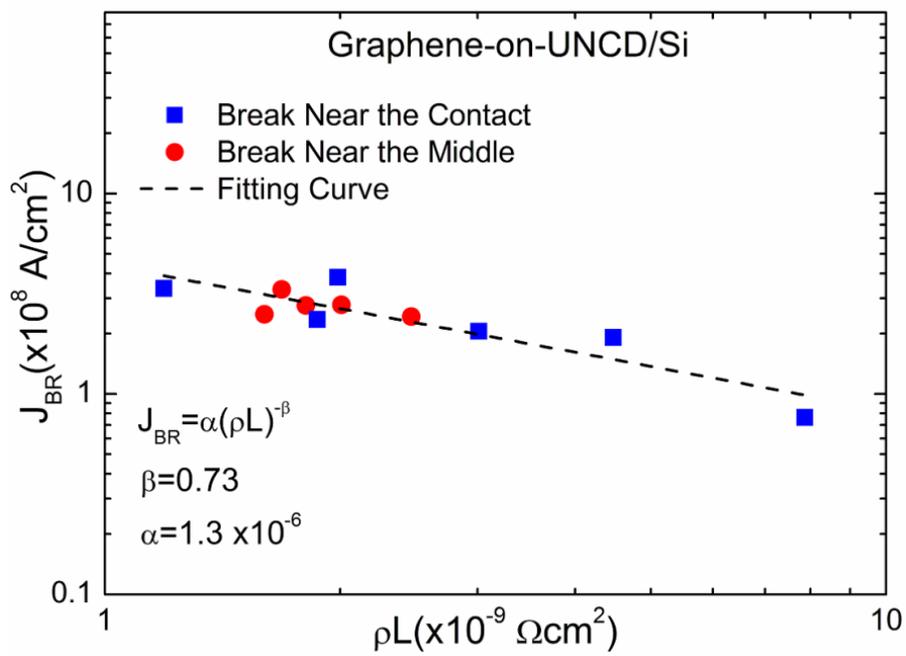

Figure 4